\centerline{\bf Admissibility of initial data in spherical collapse}
\bigskip
\centerline{H. M.  Antia}
\centerline{Tata Institute of Fundamental Research, Homi Bhabha Road, Mumbai
400 005, India
}
\centerline{email: antia@tifrvax.tifr.res.in}
\vskip 2 cm
{\narrower

\centerline{\bf Abstract}

Gravitational collapse of a spherically symmetric  cloud has been
extensively studied to investigate the nature of resulting singularity.
However, there has been considerable debate about the admissibility of
certain initial density distributions. Using the Newtonian limit of the
equations governing collapse of a fluid with an equation of state
$p=p(\rho)$ it is
shown that the density distribution has to be even function of
$r$ in a spherically symmetric situation provided $dp/d\rho\ne0$
even in comoving coordinates. We show that recent claim by Singh
that the discrepancy pointed out earlier is due to their use of comoving
coordinates is totally incorrect. It is surprising that he expects
the use of comoving coordinates to make any difference in this matter.
It is also argued that the strong curvature naked singularities in
gravitational collapse of spherically symmetric dust do not violate
the cosmic censorship hypothesis.
\bigskip
}

There has been considerable discussion about the nature of allowed initial
data in spherically symmetric collapse. Unnikrishnan [1,2]
has argued that density has to be an even function of radial distance
when a realistic equation of state with pressure is used.
On the other hand, Jhingan, Joshi and Singh [3]
have claimed that odd terms in density cannot be ruled out.
Antia [4] has pointed out that their claim is incorrect as they have
not included the continuity equation in their analysis. Recently
Singh [5] has claimed that this difference is because of their using the
Lagrangian (or comoving) coordinates. In this paper we show that
claim of Singh is totally incorrect and even in comoving coordinates
the density has to be even function of radial distance when the
matter obeys an equation of state of the form $P=P(\rho)$ with
$dP/d\rho\ne0$.

Following Singh [5]  and using his notation, we expand
$$R(t,r)=r(a_0(t)+ra_1(t)+r^2a_2(t)+\cdots) \eqno(1)$$
However, in the comoving frame as used in [3] at $t=0$ the scaling
is chosen to ensure $R(0,r)=r$, which yields
$$a_0(0)=1,\qquad a_i(0)=0\quad (i>0)\eqno(2)$$
Now according to [5] the coefficient $a_1$ satisfies the differential
equation
$${d a_1\over dt}={a_1\over a_0}{da_0\over dt}\eqno(3)$$
With the initial condition given by Eq.(2) this equation will yield
$a_1=0$ and hence the proof given by Antia [4] for the Eulerian equations
can be continued to show that all the odd terms in expansion for 
density will vanish even in the Lagrangian case. Admittedly, the proof
will be more involved in this case as compared to the Eulerian case
considered in [4]. If the scaling
$R(0,r)=r$ is not used then $a_1$ may be non-zero but in that case
$r$ will not represent radial distance.

This result can be more easily seen if the continuity
equation is written in the usual form given by eq.(3) of [3], i.e.
$$\rho(t,r)={F'(r)\over R^2R'}\eqno(4)$$
where 
$$F(r)=\int_0^r \rho(0,r)r^2\;dr\eqno(5)$$
Thus we can expand
$$F(r)=r^3(F_0+rF_1+r^2F_2+\cdots)\eqno(6)$$
Further, since $\rho_1$ the linear term in the expansion of density vanishes
it immediately follows that $F_1=0$. Now using these expansions in
Eq.(4) it follows that $a_1(t)=0$. Thus the mistake of
Singh [5] is clear, although he has differentiated the continuity
equation with respect to time $t$ he has not used the initial conditions
to obtain the solution. It is not clear why he has not used the form of
continuity equation as used by them in [3].
As a result, the claims made by Singh [5] are totally incorrect.

It is surprising that Singh thinks that the Lagrangian form of equations
will give different results as compared to the Eulerian form, since
at least for $t=0$, $R=r$ and one cannot have a different expansion
in the two frames. Further, it is clear that the continuity equation
was not used in [3], and the attempt by Singh [5] to use the
continuity equation is merely a poor after thought.

As far as instability of strong curvature naked singularities in
spherically symmetric dust collapse is concerned, Antia [4] has
already argued that the proof is valid for the general case considered
by Singh \& Joshi [6]. In fact Jhingan, Joshi and Singh [3] have
themselves shown that strong curvature naked singularities are
non-generic and arise from initial data set of measure zero.
If that is true such solutions are clearly unstable, since there would
be some allowed perturbation which will not be in the class of solutions
leading to strong curvature naked singularity.
Further, Singh [7] has stated the cosmic censorship hypothesis as
``Gravitational collapse of physically reasonable matter starting
from generic initial data leads to the formation of a black-hole,
not a naked singularity''. Thus clearly according to Singh, the
counterexamples with strong curvature naked singularities [3,6,8]
in gravitational collapse of spherically symmetric dust
do not even violate the cosmic censorship hypothesis, because these do
not arise from generic initial data!

{\bf Acknowledgments:} I would like to thank Professor Pankaj S. Joshi,
Dr. T. P. Singh and Dr. I. H. Dwivedi for a series of long
and stormy discussions.

\bigskip
{\parindent=0 pt\everypar{\hangindent=20 pt}

[1] C. S. Unnikrishnan, Gen.\ Rel.\ Grav.\ {\bf 26} (1994) 655.

[2] C. S. Unnikrishnan, Phys.\ Rev.\ {\bf D53} (1996) R580.

[3] S. Jhingan, P. S. Joshi and T. P. Singh, Class.\ Quantum Grav.\ 
{\bf 13} (1996) 3057.

[4] H. M. Antia, gr-qc/9701023.

[5] T. P. Singh, gr-qc/9702023.

[6] T. P. Singh and P. S. Joshi, Class.\ Quantum Grav.\ {\bf 13}
(1996) 559.

[7] T. P. Singh, gr-qc/9606016.

[8] P. S. Joshi and T. P. Singh, Phys.\ Rev.\ {\bf D51} (1995) 6778.

}

\end